# Temporal Action-Graph Games:
# A New Representation for Dynamic Games

**Albert Xin Jiang**
Department of Computer Science
University of British Columbia
jiang@cs.ubc.ca

**Kevin Leyton-Brown**
Department of Computer Science
University of British Columbia
kevinlb@cs.ubc.ca

**Avi Pfeffer**
Charles River Analytics
apfeffer@cra.com

## Abstract

In this paper we introduce temporal action graph games (TAGGs), a novel graphical representation of imperfect-information extensive form games. We show that when a game involves anonymity or context-specific utility independencies, its encoding as a TAGG can be much more compact than its direct encoding as a multiagent influence diagram (MAID). We also show that TAGGs can be understood as indirect MAID encodings in which many deterministic chance nodes are introduced. We provide an algorithm for computing with TAGGs, and show both theoretically and empirically that our approach improves significantly on the previous state of the art.

## 1 Introduction

Game theory is the main formal model used to study decision-making in the presence of other rational agents (see e.g., [5]). In this paper we investigate game-theoretic models that include sequential moves (so-called extensive-form games) and utility uncertainty (so-called Bayesian games, or "chance nodes" in the extensive form). When agents move sequentially and are able to perfectly observe all moves, extensive-form games are said to exhibit *perfect information*; otherwise, extensive-form games exhibit *imperfect information*.

In the last decade, researchers have begun to study compact representations of various game types, inspired by the success of probabilistic graphical models. For imperfect-information extensive form games, the most influential model is multiagent influence diagrams, or MAIDs [11]. (*Game networks* are a very similar representation and were invented concurrently [13].) MAIDs are compact when players' utility functions exhibit independencies; such compactness can also be leveraged for computational benefit [2].

**Example 1:** Twenty cars are approaching a tollbooth with three lanes. The drivers must decide which lane to use. The cars arrive in four waves of five cars each. In each wave, the drivers must pick lanes simultaneously, and can see the number of cars before them in each lane. A driver's utility decreases with the number of cars that chose the same lane either before him or at the same time.

When represented as a MAID, the game of Example 1 contains very little structure, meaning that computation would be highly inefficient. However, the game really is highly structured: agents' payoffs exhibit context-specific independence (utility depends only on the number of cars in the chosen lane) and agents' payoffs exhibit anonymity (utility depends on the numbers of other agents taking given actions, not on these agents' identities). The problem with a straightforward MAID representation of this game is that it does not capture either of these kinds of payoff structure.

A wider variety of compact game representations exist for simultaneous-move (i.e., non-extensive-form) games [10]. In particular, several of these game representations can compactly represent anonymity and context-specific independence (CSI) structures (see, e.g., [20; 14]). *Action-Graph Games (AGGs)* unify these past representations by compactly representing both anonymity and CSI while still retaining the ability to represent any game [1; 9]. Furthermore, structure in AGGs can be leveraged for computational benefit [8]. AGGs have been used to model real-world systems such as auctions for search engine keywords [21]. However, AGGs are unable to represent the game presented in Example 1 because they cannot describe sequential moves or imperfect information.

In this paper we present a new representational framework called Temporal Action-Graph Games (TAGGs) that allows us to capture this kind of structure. Like AGGs, TAGGs can represent anonymity and CSI, but unlike AGGs they can also represent games with dynamics, imperfect information and uncertainty. We first define the representation of TAGGs, and then show formally how they define a game using an induced Bayesian network (BN). We demonstrate that TAGGs can represent any MAID, but can also represent situations that are hard to capture naturally as MAIDs. If the TAGG representation of a game contains anonymity or CSI, the induced BN will have special structure that can be exploited by inference algorithms. We present an algorithm for computing expected utility of TAGGs that exploits this structure. Our algorithm first transforms the induced BN to another BN that represents the structure more explicitly, then computes expected utility using a specialized inference algorithm on the transformed BN. We show that it performs better than using a MAID in which the structure is not represented explicitly, and better than using a standard BN inference algorithm on the transformed BN.

## 2 Representation

### 2.1 AGG

We first introduce the core concepts of (simultaneous-move) AGGs. An *Action-Graph Game (AGG)* is a game played by a set of agents $N = \{1, \ldots, n\}$ on a set of *action nodes* $\mathcal{A}$. To play the game, each agent $i$ simultaneously chooses an action node $A_i \in \mathcal{A}_i$, where $\mathcal{A}_i \subseteq \mathcal{A}$. Each action node thus corresponds to an action choice that is available to one or more of the agents. Once the agents have made their choices, for each action node $A \in \mathcal{A}$, an *action count* is tallied, which is the number of agents that have chosen $A$. The *action graph* is a directed graph on the action nodes $\mathcal{A}$. We say $A'$ is a neighbor of $A$ if there is an edge from $A'$ to $A$. An agent's utility depends only on the node she chose and the action counts on the neighbors of the chosen node. This allows us to represent the utilities of the game by specifying a utility function $U_A$ for each action node $A$, that maps from the set of *configurations* over the neighbors of $A$ (vectors of action counts) to a real value.

Consider a simultaneous-move version of Example 1, where all the cars arrive at the tollbooth together in one wave, and have to pick lanes simultaneously. A driver's utility depends on the number of cars that chose the same lane. This game can be compactly represented as an AGG. We have one action node corresponding to each lane of the tollbooth. Each agent can choose any of the lanes, so $\mathcal{A}_i = \mathcal{A}$ for all $i$. The only edges in the action graph are from each action node to itself, since the utility of choosing a lane only depends on the action count on the same lane.

### 2.2 TAGG

At a high level, *Temporal Action-Graph Games (TAGGs)* extend the AGG representation by introducing the concepts of *time*, *uncertainty* and *imperfect information*, while adapting the AGG concepts of action nodes and action-specific utility functions to the dynamic setting. We first give an informal description of these concepts.

**Temporal structure.** A TAGG describes a dynamic game played over a series of time steps $1, \ldots, T$, on a set of action nodes $\mathcal{A}$. At each time step a version of a static AGG is played by a subset of agents on $\mathcal{A}$, and the action counts on the action nodes are accumulated.

**Chance variables.** TAGGs model uncertainty via *chance variables*. Like random variables in a Bayes net, a chance variable is associated with a set of parents and a conditional probability table (CPT). The parents may be action nodes or other chance variables. Each chance variable is associated with an instantiation time; once instantiated, its value stays the same for the rest of the game.

**Decisions.** At each time step one or more agents move simultaneously, represented by agent-specific *decisions*. TAGGs model imperfect information by allowing each agent to condition his decision on observed values of a given subset of decisions, chance variables, and the previous time step's action counts.

**Action nodes.** Each decision is a choice of one from a number of available *action nodes*. As in AGGs, the same action may be available to more than one player. Action nodes provide a time-dependent tally: the *action count* for each action $A$ in each time step $\tau$ is the number of times $A$ has been chosen during the time period $1, \ldots, \tau$.

**Utility functions.** There is a *utility function* $U_A^\tau$ associated with each action $A$ at each time $\tau$, which specifies the utility a player receives at time $\tau$ for having chosen action $A$. Each $U_A^\tau$ has a set of parents which must be action nodes or chance variables. The utility of playing action $A$ depends only on what happens over these parents. An agent who took action $A$ (once) may receive utility at multiple times (e.g., short-term cost and long-term benefit); this is captured by associating a set of payoff times with each decision. An agent's overall utility is defined as the sum of the utilities received at all time steps.

Play of a TAGG can be summarized as follows:

1. At time 0, action counts are initialized to zero; chance variables with instantiation time 0 are instantiated,

2. At each time $\tau \in \{1, \ldots, T\}$:
   (a) all agents with decisions at $\tau$ observe the appropriate action counts, chance variables, and decisions, if any.
   (b) all decisions at $\tau$ are made simultaneously.
   (c) action counts at $\tau$ are tallied.
   (d) chance variables at $\tau$ are instantiated.
   (e) for each action $A$, utility function $U_A^\tau$ is evaluated, with this amount of utility accruing to every agent who took action $a$ at a decision whose payoff times include $\tau$; the result is not revealed to any of the players.[1]

---

[1]If an agent plays action $A$ for two decisions that have the

3. At the end of the game, each agent receives the sum of all utility allocations throughout the game.

Intuitively, the process can be seen as a sequence of AGGs played over time. At each time step $\tau$, the players that have a decision at time $\tau$ participate in a simultaneous-move AGG on the set of action nodes, whose action counts are initialized to be the counts at $\tau - 1$. Each action $A$'s utility function is $U_A^\tau$ and $A$'s neighbors in the action graph correspond to the parents of $U_A^\tau$. For the tollbooth games of Example 1, the TAGG has actions corresponding to lanes, and the action graphs are the same across different times, containing only self edges from each action to itself.

Before formally defining TAGGs, we need to first define the concept of a *configuration* at time $\tau$ over a set of action nodes, decisions and chance variables, which is intuitively an instantiation at time $\tau$ of a corresponding set of variables.

**Definition 2:** Given a set of action nodes $\mathcal{A}$, a set of decisions $\mathcal{D}$, a set of chance variables $\mathcal{X}$, and a set $B \subseteq \mathcal{A} \cup \mathcal{X} \cup \mathcal{D}$, a *configuration* at time $\tau$ over $B$, denoted as $C_B^\tau$, is a $|B|$-tuple of values, one for each node in $B$. For each node $b \in B$, the corresponding element in $C_B^\tau$, denoted as $C^\tau(b)$, must satisfy the following:

- if $b \in \mathcal{A}$, $C^\tau(b)$ is an integer in $\{0, \ldots, |\mathcal{D}|\}$ specifying the action count on $b$ at $\tau$, i.e. the number of times action $b$ has been chosen during the time period $1, \ldots, \tau$.
- if $b \in \mathcal{D}$, $C^\tau(b)$ is an action in $\mathcal{A}$, specifying the action chosen at $D$.
- if $b \in \mathcal{X}$, $C^\tau(b)$ is a value from the domain of the random variable, $Dom[b]$.

Let $\mathcal{C}_B^\tau$ be the set of all configurations at $\tau$ over $B$.

We now offer formal definitions of chance variables, decisions, and utility functions.

**Definition 3:** A *chance variable* $X$ is defined by:

1. a domain $\text{Dom}[X]$, which is a nonempty finite set;
2. a set of parents $\text{Pa}[X]$, which are a set of chance variables and/or actions;
3. an instantiation time $t(X)$, which specifies the time at which the action counts in $\text{Pa}[X]$ are instantiated;
4. a CPT $\Pr(X|\text{Pa}[X])$, which specifies the conditional probability distribution of $X$ given each configuration $C_{\text{Pa}[X]}^{t(X)}$.

same payoff time $\tau$, then the agent receives twice the value of $U_A^\tau$.

We require that each chance variable's instantiation time be no earlier than its parent chance variable's instantiation times, i.e. if chance variable $X' \in \text{Pa}[X]$, then $t(X') \leq t(X)$.

**Definition 4:** A *decision* $D$ is defined by:

1. the player making the decision, $\text{pl}(D)$. A player may make multiple decisions, but not more than one at the same time step. The set of decisions belonging to a player $\ell$ is denoted by $\text{Decs}[\ell]$.
2. its *decision time* $t(D) \in \{1, \ldots, T\}$.
3. its *action set* $\text{Dom}[D]$, a nonempty set of actions.
4. the set of *payoff times* $\text{pt}(D) \subseteq \{1, \ldots, T\}$. We assume that $\tau \geq t(D)$ for all $\tau \in \text{pt}(D)$.
5. its *observation set* $O[D]$: a set of decisions, actions, and chance variables, whose configuration at time $t(D) - 1$ (i.e. $C_{O[D]}^{t(D)-1}$) is observed by $\text{pl}(D)$ prior to making the decision. We require that if decision $D'$ is an observation of $D$, then $t(D') < t(D)$. Furthermore if chance variable $X$ is an observation of $D$, then $t(X) < t(D)$.

**Definition 5:** Each action $A$ at each time $\tau$ is associated with one *utility function* $U_A^\tau$. Each $U_A^\tau$ is associated with a set of parents $\text{Pa}[U_A^\tau]$, which is a set of actions and chance variables. We require that if chance variable $X \in \text{Pa}[U_A^\tau]$, then $t(X) \leq \tau$. Each utility function $U_A^\tau$ is a mapping from the set of configurations $\mathcal{C}_{\text{Pa}[U_A^\tau]}^\tau$ to a real value.

We can now formally define TAGGs.

**Definition 6:** A Temporal Action-Graph Game (TAGG) is a tuple $(N, T, \mathcal{A}, \mathcal{X}, \mathcal{D}, \mathcal{U})$, where:

1. $N = \{1, \ldots, n\}$ is a set of *players*.
2. $T$ is the *duration* of the game.
3. $\mathcal{A}$ is a set of *actions*.
4. $\mathcal{X}$ is a set of *chance variables*. Let $\mathcal{G}$ be the induced directed graph over $\mathcal{X}$. We require that $\mathcal{G}$ be a directed acyclic graph (DAG).
5. $\mathcal{D}$ is the set of *decisions*. we require that each decision $D$'s action set $\text{Dom}[D] \subseteq \mathcal{A}$.
6. $\mathcal{U} = \{U_A^\tau : A \in \mathcal{A}, 1 \leq \tau \leq T\}$ is the set of *utility functions*.

First, let us see how to represent Example 1 as a TAGG. The set $N$ corresponds to the cars. The duration $T = 4$. We have one action node for each lane. For each time $\tau$, we have five decisions, each belonging to a car that arrives at time $\tau$. The action set for each decision is the entire set $\mathcal{A}$. The payoff time for each decision is the time the decision is made, i.e., $pt(D) = \{t(D)\}$. Each decision has all actions as observations. For each $A$ and $\tau$, the utility $U_A^\tau$ has $A$ as its only parent. The representation size of each utility function is at most $n$; the size of the entire TAGG is $O(|\mathcal{A}|Tn)$.

The TAGG representation is useful beyond compactly representing MAIDs. The representation can also be used to specify information structures that would be difficult to represent in a MAID. For example, we can represent games in which agents' abilities to observe the decisions made by previous agents depend on what actions these agents took.

**Example 7:** There are $2T$ ice cream vendors, each of which must choose a location along a beach. For every day from 1 to $T$, two of the vendors simultaneously set up their ice cream stands. Each vendor lives in one of the locations. When a vendor chooses an action, it knows the location of vendors who set up stands in previous days in the location where it lives or in one of the neighboring locations. The payoff to a vendor in a given day depends on how many vendors set up stands in the same location or in a neighboring location.

Example 7 can be represented as a TAGG, the key elements of which are as follows. There is an action $A$ for each location. Each player $j$ has one decision $D_j$, whose observations include actions for the location $j$ lives in and neighboring locations. The payoff time for each decision is $T$, and the utility function $U_A^T$ has $A$ and its neighboring locations as parents.

Let us consider the size of a TAGG. It follows from Definition 6 that the space bottlenecks of the representation are the CPTs $\Pr(X|\text{Pa}[X])$ and the utility functions $U_A^\tau$, which have polynomial sizes when the numbers of their parents are bounded by a constant.

**Lemma 8:** *Given TAGG $(N, T, \mathcal{A}, \mathcal{X}, \mathcal{D}, \mathcal{U})$, if $\max_{X \in \mathcal{X}} |Pa[X]|$ and $\max_{U \in \mathcal{U}} |Pa[U]|$ are bounded by a constant, then the size of the TAGG is bounded by a polynomial in $\max_{X \in \mathcal{X}} Dom[X]$, $|\mathcal{X}|$, $|\mathcal{D}|$, $|\mathcal{U}|$, and $T$.*

### 2.3 Strategies

We now define an agent's strategies in a TAGG. We start with *pure strategies*, where at each decision $D$, an action is chosen deterministically as a function of observed information, i.e., the configuration $C_{O[D]}^{t(D)-1}$. In game theory, pure strategies are not generally sufficient (e.g., for guaranteeing existence of Nash equilibrium). We often want to consider strategies in which players randomize over their choices when making decisions. A *mixed strategy* of a player $i$ is a probability distribution over pure strategies of $i$. Use of mixed strategies is problematic, because the choices at different decisions may be correlated, which allows a player to condition her later choice on her earlier decisions, regardless of whether these earlier decisions are observed by her. In other words, mixed strategies allow a player to act as if she observes more information than what is specified by the game. We thus restrict our attention to *behavior strategies*, in which the action choices at different decisions are randomized independently.

**Definition 9:** A behavior strategy at decision $D$ is a function $\sigma^D : \mathcal{C}_{O[D]}^{t(D)-1} \to \varphi(\text{Dom}[D])$, where $\varphi(\text{Dom}[D])$ is the set of probability distributions over $\text{Dom}[D]$. A behavior strategy for player $i$, denoted $\sigma_i$, is a tuple consisting of a behavior strategy for each of her decisions. A behavior strategy profile $\boldsymbol{\sigma} = (\sigma_1, \ldots, \sigma_n)$ consists of a behavior strategy $\sigma_i$ for all $i$.

An agent has *perfect recall* when she never forgets her action choices and observations at earlier decisions. Equilibria in behavior strategies always exist in games of perfect recall [12]. However, strategies in perfect recall games can be computationally expensive to represent and reason about. In single-agent settings, perfect recall can be relaxed using limited memory influence diagrams (LIMIDs) [17]. For multi-agent imperfect recall games, existence of Nash equilibria in behavior strategies is not guaranteed; nevertheless one might still want to search for Nash equilibria using heuristic algorithms. There have also been positive results for certain types of imperfect recall games, where information irrelevant to the utility can be safely forgotten, e.g. [15]. The TAGG representation does not enforce perfect recall; TAGGs can represent perfect recall games as well as non-perfect-recall games.

### 2.4 Expected Utility

Now we use the language of Bayesian networks to formally define an agent's expected utility in a TAGG given a behavior strategy profile $\boldsymbol{\sigma}$. Specifically, we define an *induced BN* that formally describes how the TAGG is played out. The induced BN is over a set of random variables representing decisions, chance variables, action counts and utilities. Given a behavioral strategy profile, decisions, chance variables and utilities can naturally be understood as random variables. On the other hand, action counts are time-dependent. Thus, we have a separate action count variable for each action at each time step.

**Definition 10:** Let $A \in \mathcal{A}$ be an action and $\tau \in \{1, ..., T\}$ be a time point. $A^\tau$ denotes the *action count variable* representing the number of times $A$ was chosen from time 1 to time $\tau$. Let $A^0$ be the variable which is constantly 0.

We would like to define expected utility for each player, which is the sum of expected utilities of her decisions. On the other hand, the utility functions in TAGGs are action specific. To bridge the gap, we create new decision-payoff variables in the induced BN that represent the utilities of decisions received at each of their payoff time points.

**Definition 11:** Given a TAGG and a behavior strategy profile $\boldsymbol{\sigma}$, the *induced BN* is defined over the following *variables*: for each decision $D \in \mathcal{D}$ there is a variable which by abuse of notation we shall also denote by $D$; for each chance variable $X \in \mathcal{X}$ there is a variable which we shall also denote by $X$; there is a variable $A^\tau$ for each action $A \in \mathcal{A}$ and time step $\tau \in \{1,...,T\}$; for each utility function $U_A^\tau$ for actions $A \in \mathcal{A}$ and time points $\tau \in \{1,...,T\}$, there is a utility variable also denoted by $U_A^\tau$; for each decision $D$ and each time $\tau \in \text{pt}(D)$, there is a decision-payoff variable $u_D^\tau$.

We define the *actual parents* of each variable $V$, denoted $\text{APa}[V]$, as follows: The actual parents of a decision variable $D$ are the variables corresponding to $O[D]$, with each action $A_k \in O[D]$ replaced by $A_k^{t(D)-1}$. The actual parents of an action count variable $A^\tau$ are all decision variables $D$ whose decision time $t(D) \leq \tau$ and $A \in \text{Dom}[D]$. The actual parents of a chance variable $X$ are the variables corresponding to $\text{Pa}[X]$, with each action $A_k \in \text{Pa}[X]$ replaced by $A_k^{t(X)}$. The actual parents of a utility variable $U_A^\tau$ are the variables corresponding to $\text{Pa}[U_A^\tau]$, with each action $A_k \in \text{Pa}[U_A^\tau]$ replaced by $A_k^\tau$. where $\{A_1,...,A_\ell\} = \text{Dom}[D]$.

The CPDs of chance variables are the CPDs of the corresponding chance variables in the TAGG. The CPD of each decision variable $D$ is the behavior strategy $\sigma^D$. The CPD of each utility variable $U_A^\tau$ is a deterministic function defined by the corresponding utility function $U_A^\tau$. The CPD of each action count variable $A^\tau$ is a deterministic function that counts the number of decisions in $\text{APa}[A]$ that are assigned value $A$. The CPD of each decision-payoff variable $u_D^\tau$ is a *multiplexer*, i.e. a deterministic function that selects the value of its utility variable parent according to the choice of its decision parent. For example, if the value of $D$ is $A_k$, then the value of $u_D^\tau$ is the value of $U_{A_k}^\tau$.

**Theorem 12:** *Given a TAGG, let $\mathcal{F}$ be the directed graph over the variables of the induced BN in which there is an edge from $V_1$ to $V_2$ iff $V_1$ is an actual parent of $V_2$. Then $\mathcal{F}$ is acyclic.*

This follows from the definition of TAGGs and the way we set up the actual parents in Definition 11.

By Theorem 12, the induced BN defines a joint probability distribution over its variables, which we denote by $P^{\boldsymbol{\sigma}}$. Given $\boldsymbol{\sigma}$, denote by $E^{\boldsymbol{\sigma}}[V]$ the expected value of variable $V$ in the induced BN. We are now ready to define the expected utility to players under behavior strategy profiles.

**Definition 13 :** The expected utility to player $\ell$ under behavior strategy profile $\boldsymbol{\sigma}$ is $\text{EU}^{\boldsymbol{\sigma}}(\ell) = \sum_{D \in \text{Decs}[\ell]} \sum_{\tau \in \text{pt}(D)} E^{\boldsymbol{\sigma}}[u_D^\tau]$.

Figure 1 shows an induced BN of a TAGG based on Example 1 with six cars and three lanes. Note that although we use squares to represent decision variables, they are random variables and not actual decisions as in influence diagrams.

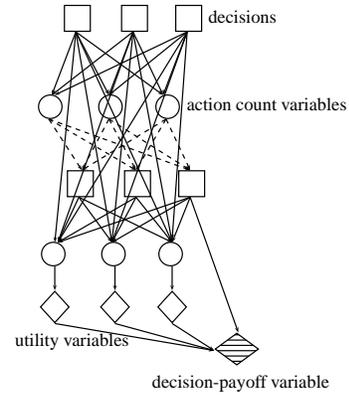

Figure 1: Induced BN of the TAGG of Example 1, with 2 time steps, 3 lanes, and 3 players per time step. Squares represent decision variables, circles represent action count variables, diamonds represent utility variables and shaded diamonds represent decision-payoff variables. To avoid cluttering the graph, we only show utility variables at time step 2 and a decision-payoff variable for one of the decisions.

### 2.5 The Induced MAID of a TAGG

Given a TAGG we can construct a MAID that describes the same game. We use a similar construction as the induced Bayesian Network, but with two differences. First, instead of decision variables with CPDs assigned by $\boldsymbol{\sigma}$, we have decision nodes in the MAID. Second, each decision-payoff variable $u_D^\tau$ becomes a utility node for player $\text{pl}(D)$ in the MAID. The resulting MAID describes the same game as the TAGG, because it offers agents the same strategies and their expected utilities are defined by the same BN. We call this the *induced MAID* of the TAGG.

### 2.6 Expressiveness

It is natural to ask about the *expressiveness* of TAGGs: what games can we represent? It turns out that TAGGs are at least as expressive as MAIDs.

**Lemma 14:** *Any MAID can be represented as a TAGG with the same space complexity.*

As a result, TAGGs can represent any extensive form games representable as MAIDs. These include all perfect recall games, and the subclass of imperfect recall games where each information set does not involve multiple time steps.

On the other hand, since the induced MAID of a TAGG is payoff equivalent to the TAGG, it trivially follows that any TAGG can be represented by a MAID. However, the induced MAID has a large in-degree, and can thus be exponentially larger than the TAGG. For example, in the games of Examples 1 and 7, the induced MAIDs have max in-degrees that are equal to the number of decisions, which implies that the sizes of the MAIDs grow exponentially with the number of decisions, whereas the sizes of the TAGGs for the same games grow linearly in the number of decisions. This is not surprising, since TAGGs can exploit more kinds of structure in the game (CSI, anonymity) compared to a straightforward MAID representation. In Section 3.1 we show that the induced MAID can be transformed into a MAID that explicitly represents the underlying structure. The size of the transformed MAID is polynomial in the size of the TAGG.

The TAGG representation is also a true generalization of AGGs, since any simultaneous-move AGG can be straightforwardly represented as a TAGG with $T = 1$.

## 3 Computing Expected Utility

A compact game representation is not very useful if we cannot perform game-theoretic computations efficiently with respect to the size of the representation. In this section, we consider the task of computing expected utility $EU^{\sigma}[j]$ to a player $j$ given a mixed strategy profile $\sigma$. Computation of EU is an essential step in many game-theoretic computations, such as finding a best response given other players' strategy profile, and the iterated best response algorithm, which is not guaranteed to converge but finds a Nash equilibrium if it converges. In Section 4 we discuss extending our methods in this section to a subtask in the Govidan-Wilson algorithm for computing Nash equilibria, which is guaranteed to converge. Thus we can achieve speedup of all these computations by speeding up their bottleneck steps.

One benefit of formally defining EU in terms of BNs is that now the problem of computing EU can be naturally cast as a Bayes-net inference problem. By Definition 13, $EU^{\sigma}[j]$ is the sum of a polynomial number of terms of the form $E^{\sigma}[u_D^{\tau}]$. We thus focus on computing one such $E^{\sigma}[u_D^{\tau}]$. This can be computed by applying a standard Bayes-net inference algorithm on the induced BN. In fact, Bayes-net inference is the standard approach for computing expected utility in MAIDs [11]. Thus the above approach for TAGGs is computationally equivalent to the standard approach for a natural MAID representation of the same game. In this section, we show that the induced BNs of TAGGs have special structure that can be exploited to speed up computation, and present an algorithm that exploits this structure.

### 3.1 Exploiting causal independence

The standard BN inference approach for computing EU does not take advantage of some kinds of TAGG structure. In particular, recall that in the induced network, each action count variable $A^\tau$'s parents are all previous decisions that have $A^\tau$ in their action sets, implying large in-degrees for action variables. Considering for example the clique-tree algorithm, this means large clique sizes, which is problematic because running time scales exponentially in the largest clique size of the clique tree. However, the CPDs of these action count variables are structured counting functions. Such structure is an instance of *causal independence* in BNs [7]. It also corresponds to anonymity structure for static game representations like symmetric games and AGGs [8].

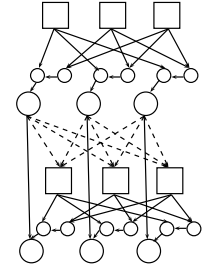

Figure 2: The transformed BN of the tollbooth game from Figure 1 with 3 lanes and 3 cars per time step.

We can exploit this structure to speed up computation of expected utility in TAGGs. Our approach is a specialization of Heckerman and Breese's method [7] for exploiting causal independence in BNs. At a high level, Heckerman and Breese's method transforms the original BN by creating new nodes that represent intermediate results, and re-wiring some of the arcs, resulting in an equivalent BN with small in-degree. They then apply conventional inference algorithms on the new BN. For example, given an action count variable $A_k^\tau$ with parents $\{D_1 \ldots D_\ell\}$, create a node $M_i$ for each $i \in \{1 \ldots \ell - 1\}$, representing the count induced by $D_1 \ldots D_i$. Then, instead of having $D_1 \ldots D_\ell$ as parents of $A_k^\tau$, its parents become $D_\ell$ and $M_{\ell-1}$, and each $M_i$'s parents are $D_i$ and $M_{i-1}$. The resulting graph would have in-degree at most 2 for $A_k^\tau$ and the $M_i$'s.

In our induced BN, the action count variables $A_k^t$ at earlier time steps $t < \tau$ already represent some of these intermediate counts, so we do not need to duplicate them. Formally, we modify the original BN in the following way: for each action count variable $A_k^\tau$, first remove the edges from its current parents. Instead, $A_k^\tau$ now has two parents: the action count variable $A_k^{\tau-1}$ and a new node $M_{A_k}^\tau$ representing the contribution of decisions at time $\tau$ to the count of $A_k$. If there is more than one decision at time $\tau$ that has $A_k$ in its action set, we create intermediate variables as in Heckerman and Breese's method. We call the resulting BN the *transformed BN* of the TAGG. Figure 2 shows the transformed BN of the tollbooth game whose induced BN was given in Figure 1.

We can then use standard algorithms to compute probabilities $P(u_D^{t'})$ on the transformed BN. For classes of BNs with bounded treewidths, these probabilities (and thus $E[u_D^{t'}]$) can be computed in polynomial time.

### 3.2 Exploiting temporal structure

In practice, the standard inference approaches use heuristics to find an elimination ordering. This might not be optimal for our BNs. We present an algorithm based on the idea of eliminating variables in the temporal order. For the rest of the section, we fix $D$ and a time $t' \in \text{pt}(D)$ and consider the computation of $E^{\boldsymbol{\sigma}}[u_D^{t'}]$.

We first group the variables of the induced network by time steps: variables at time $\tau$ include decisions at $\tau$, action count variables $A^\tau$, chance variables $X$ with instantiation time $\tau$, intermediate nodes between decisions and action counts at $\tau$, and utility variables $U_A^\tau$. As we are only concerned about $E^{\boldsymbol{\sigma}}[u_D^{t'}]$ for a $t' \in \text{pt}(D)$, we can safely discard the variables after time $t'$, as well as utility variables before $t'$. It is straightforward to verify that the actual parents of variables at time $\tau$ are either at $\tau$ or before $\tau$.

We say a network satisfies the *Markov property* if the actual parents of variables at time $\tau$ are either at $\tau$ or at $\tau-1$. Parts of the induced BN (e.g. the action count variables) already satisfy the Markov property, but in general the network does not satisfy the property. Exceptions include chance variable parents and decision parents from more than one time step ago.

Given an induced BN, we can transform it into an equivalent network satisfying the Markov property. If a variable $V_1$ at $t_1$ is a parent of variable $V_2$ at $t_2$, with $t_2 - t_1 > 1$, then for each $t_1 < \tau < t_2$ we create a dummy variable $V_1^\tau$ belonging to time $\tau$ so that we copy the value of $V_1$ to $V_1^{t_2-1}$. We then delete the edge from $V_1$ to $V_2$ and add an edge from $V_1^{t_2-1}$ to $V_2$.

The Markov property is computationally desirable because variables in time $\tau$ d-separate past variables from future variables. A straightforward approach to exploiting the Markov property is the following: as $\tau$ goes from 1 to $t'$, compute the joint distribution over variables at $\tau$ using the joint distribution over variables at $\tau - 1$.

In fact, we can do better by adapting the *interface algorithm* [4] for dynamic Bayesian networks (DBNs) to our setting.[2] Define the *interface* $\mathbf{I}^\tau$ to be the set of variables in time $\tau$ that have children in time $\tau + 1$. $\mathbf{I}^\tau$ d-separates *past* from *future*, where *past* is all variables before $\tau$ and non-interface variables in $\tau$, and *future* is all variables after $\tau$.

In an induced BN, $\mathbf{I}^\tau$ consists of: action count variables at time $\tau$; chance variables $X$ at time $\tau$ that have children in *future*; decisions at $\tau$ that are observed by *future* decisions; decision $D$ which is a parent of $u_D^{t'}$, and dummy variables created by the transform.

We define the set of *effective variables* at time $\tau$, denoted by $\mathbf{V}^\tau$, as the subset of $\mathbf{I}^\tau$ that are ancestors of $u_D^{t'}$. For time $t'$, we let $\mathbf{V}^{t'} = \{u_D^{t'}\}$. Intuitively, at each time step $\tau$ we only need to keep track of the distribution $P(\mathbf{V}^\tau)$, which acts as a sufficient statistic as we go forward in time. For each $\tau$, we calculate $P(\mathbf{V}^\tau)$ by conditioning on instantiations of $P(\mathbf{V}^{\tau-1})$. The interface algorithm for TAGGs can be summarized as the following:

1. compute distribution $P(\mathbf{V}^0)$
2. for $\tau = 1$ to $t'$
   (a) for each instantiation of $\mathbf{V}^{\tau-1}$, $v_j^{\tau-1}$, compute the distribution over $\mathbf{V}^\tau$: $P\left(\mathbf{V}^\tau | \mathbf{V}^{\tau-1} = v_j^{\tau-1}\right)$
   (b) $P(\mathbf{V}^\tau) = \sum_v P\left(\mathbf{V}^\tau | \mathbf{V}^{\tau-1} = v\right) P\left(\mathbf{V}^{\tau-1} = v\right)$
3. since $\mathbf{V}^{t'} = \{u_D^{t'}\}$, we now have $P(u_D^{t'})$
4. return the expected value $E[u_D^{t'}]$

We can further improve on this, in particular on the subtask of computing $P(\mathbf{V}^\tau | \mathbf{V}^{\tau-1})$. We observe that there is also a temporal order among variables in each time $\tau$: first the decisions and intermediate variables, then action count variables, and finally chance variables. Partition $\mathbf{V}^\tau$ into four subsets consisting of action count variables $\mathbf{A}^\tau$, chance variables $\mathbf{X}^\tau$, decision variables $\mathbf{D}^\tau$ and dummy copy variables $\mathbf{C}^\tau$. Then $P(\mathbf{V}^\tau | \mathbf{V}^{\tau-1})$ can be factored into $P(\mathbf{C}^\tau | \mathbf{V}^{\tau-1}) P(\mathbf{D}^\tau, \mathbf{A}^\tau | \mathbf{V}^{\tau-1}) P(\mathbf{X}^\tau | \mathbf{A}^\tau, \mathbf{V}^{\tau-1})$. This allows us to first focus on decisions and action count variables to compute $P(\mathbf{D}^\tau, \mathbf{A}^\tau | \mathbf{V}^{\tau-1})$ and then carry out inference on the chance variables.

Calculating $P(\mathbf{D}^\tau, \mathbf{A}^\tau | \mathbf{V}^{\tau-1})$ involves eliminating all decision variables not in $\mathbf{D}^\tau$ as well as the intermediate variables. Note that conditioned on $\mathbf{V}^{\tau-1}$, all decisions at time $\tau$ are independent. This allows us to efficiently eliminate variables along the chains of intermediate variables. Let the decisions at time $\tau$ be $\{D_1^\tau, \ldots, D_\ell^\tau\}$. Let $\mathbf{M}^\tau$ be the set of intermediate variables corresponding to action count variables in $\mathbf{A}^\tau$. Let $\mathbf{M}_k^\tau$ be the subset of $\mathbf{M}^\tau$ that summarizes the contribution of $D_1^\tau, \ldots, D_k^\tau$. We eliminate variables in the order $D_1^\tau, D_2^\tau, \mathbf{M}_2^\tau, D_3^\tau, \mathbf{M}_3^\tau, \ldots, \mathbf{M}_\ell^\tau$, except for decisions in $\mathbf{D}^\tau$. The tables in the variable elimination algorithm need to keep track of at most $|\mathbf{D}^\tau| + |\mathbf{A}^\tau|$ variables. Thus the complexity of computing $P(\mathbf{D}^\tau, \mathbf{A}^\tau | \mathbf{V}^{\tau-1})$ for an instantiation of $\mathbf{V}^{\tau-1}$ is exponential only in $|\mathbf{D}^\tau| + |\mathbf{A}^\tau|$.

Computing $P(\mathbf{X}^\tau | \mathbf{A}^\tau, \mathbf{V}^{\tau-1})$ for each instantiation of $\mathbf{A}^\tau, \mathbf{V}^{\tau-1}$ involves eliminating the chance variables not in $\mathbf{X}^\tau$. Any standard inference algorithm can be applied here. The complexity is exponential in the treewidth of the induced BN restricted on all chance variables at time $\tau$, which we denote by $G^\tau$.

---

[2] Whereas in DBNs the set of variables for each time step remains the same, for our setting this is no longer the case. It turns out that the interface algorithm can be adapted to work on our transformed BNs. Also, the transformed BNs of TAGGs have more structure than DBNs, particularly within the same time step, which we exploit for further computational speedup.

Putting everything together, the bottleneck of our algorithm is constructing the tables for the joint distributions on $\mathbf{V}^\tau$, as well as doing inference on $G^\tau$.

**Theorem 15:** *Given a TAGG and behavior strategy profile $\boldsymbol{\sigma}$, if for all $\tau$, both $|\mathbf{V}^\tau|$ and the treewidth of $G^\tau$ are bounded by a constant, then for any player $j$ the expected utility $EU^{\boldsymbol{\sigma}}[j]$ can be computed in time polynomial in the size of the TAGG representation and the size of $\boldsymbol{\sigma}$.*

Our algorithm is especially effective for induced networks that are close to having the Markov property, in which case we only add a small number of dummy copy variables to $\mathbf{V}^\tau$. The time complexity of computing expected utility then grows linearly in the duration of the game. On the other hand, for induced networks far from having the Markov property, $|\mathbf{V}^\tau|$ can grow linearly as $\tau$ increases, implying that the time complexity is exponential.

### 3.3 Context-specific independence

TAGGs have action-specific utility functions, which allows them to express context-specific payoff independence: which utility function is used depends on which action is chosen at the decision. This is translated to context-specific independence structure in the induced BN, specifically in the CPD of $u_D^\tau$. Conditioned on the value of $D$, $u_D^\tau$ only depends on one of its utility variable parents.

There are several ways of exploiting such structure computationally, including conditioning on the value of the decision $D$ [3], or exploiting the context-specific independence in a variable elimination algorithm [19]. One particularly simple approach that works for multiplexer utility nodes is to decompose the utility into a sum of utilities [18]. For each utility node parent $U_k^t$ of $u_D^t$, there is a utility function $u_{D,k}^t$ that depends on $U_k^t$ and $D$. If $D = k$, $u_{D,k}^t$ is equal to $U_k^t$. Otherwise, $u_{D,k}^t$ is 0. It is easy to see that $u_D^t(U_1^t, \ldots, U_m^t, D) = \sum_{k=1}^m u_{D,k}^t(U_k^t, D)$.

We can then modify our algorithm to compute each $E[u_{D,k}^t]$ instead of $E[u_D^t]$. This results in a reduction in the set of effective variables $\mathbf{V}_k^\tau$, which are now the variables at $\tau$ that are ancestors of $u_{D,k}^t$. Furthermore, whenever $\mathbf{V}_k^\tau = \mathbf{V}_{k'}^\tau$ for some $k, k'$, the distributions over them are identical and thus can be reused.

For static games represented as TAGGs with $T = 1$, our algorithm is equivalent to the polynomial-time expected utility algorithm [8] for AGGs.

Applying our algorithm to tollbooth games of Example 1 and ice cream games of Example 7, we observe that for both cases $\mathbf{V}^\tau$ consists of a subset of action count variables at $\tau$ plus the decision whose utility we are computing. Therefore the expected utilities of these games can be computed in polynomial time if $|\mathcal{A}|$ is bounded by a constant.

## 4 Computing Nash Equilibria

Nash equilibrium is one of the central solution concepts of game theory. Since the induced MAID of a TAGG is payoff equivalent to the TAGG, algorithms for computing the Nash equilibria of MAIDs [11; 15; 2] can be directly applied to an induced MAID to find Nash equilibria of a TAGG. However, this approach does not exploit all TAGG structure. We can do better by constructing a transformed MAID, in a manner similar to the transformed BN, exploiting causal independence and CSI as in Sections 3.1 and 3.3.

We can do better yet and exploit the temporal structure as described in Section 3.2, if we use a solution algorithm that requires computation of probabilities and expected utilities. Govindan and Wilson [6] presented an algorithm for computing equilibria in perfect-recall extensive-form games using a continuation method. Blum, Shelton and Koller [2] adapted this algorithm to MAIDs. A key step in the algorithm is, for each pair of players $i$ and $j$, and one of $i$'s utility nodes, computing the marginal distribution over $i$'s decisions and their parents, $j$'s decisions and their parents, and the utility node. Our algorithm in Section 3.2 can be adapted to compute this distribution. This approach is efficient if each player only has a small number of decisions, as in the games in Examples 1 and 7.

## 5 Experiments

We have implemented our algorithm for computing expected utility in TAGGs, and run experiments on the efficiency and scalability of our algorithm. We compared three approaches for computing expected utility given a TAGG:

**Approach 1** applying the standard clique tree algorithm (as implemented by the Bayes Net Toolbox [16]) on the induced BN;

**Approach 2** applying the same clique tree algorithm on the transformed BN;

**Approach 3** our proposed algorithm.

All approaches were implemented in MATLAB. All our experiments were performed using a computer cluster consisting of machines with dual Intel Xeon 3.2GHz CPUs, 2MB cache and 2GB RAM.

We ran experiments on tollbooth game instances of varying sizes. For each game instance we measured the CPU times for computing expected utility of 100 random behavior strategy profiles. Figure 3 (left) shows the results in log scale for toll booth games with 3 lanes and 5 cars per time step, with the duration varying from 1 to 15. Approach 1 ran out of memory for games with more than 1 time step. Approach 2 was more scalable; but ran out of memory for games with more than 5 time steps. Approach 3 was the most scalable. On smaller instances it was faster than the other two approaches by an order of magnitude, and it did not run out of memory as we increased the size of the TAGGs to at least 20 time steps. For the toll booth game with 14 time steps it took 1279 seconds, which is approximately the time Approach 2 took for the game instance with 5 time steps. Figure 3 (middle) shows the results in log scale for tollbooth games with 3 time steps and 3 lanes, varying the number of cars per time step from 1 to 20. Approach 1 ran out of memory for games with more than 3 cars per time step; Approach 2 ran out of memory for games with more than 6 cars per time step; and again Approach 3 was the most scalable.

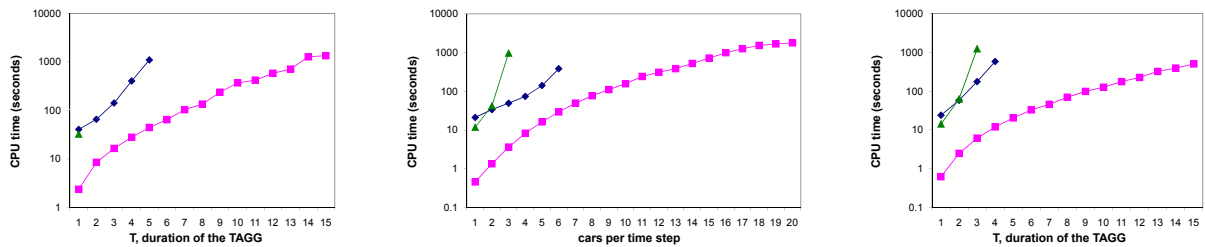

Figure 3: Running times for expected utility computation. Triangle data points represent Approach 1 (induced BN), diamonds represent Approach 2 (transformed BN), squares represent Approach 3 (proposed algorithm).

We also ran experiments on the ice cream games of Example 7. Figure 3 (right) shows the results in log scale for ice cream games with 4 locations, two vendors per time step, and durations varying from 1 to 15. The home locations for each vendor were generated randomly. Approaches 1 and 2 ran out of memory for games with more than 3 and 4 time steps, respectively. Approach 3 finished for games with 15 time steps in about the same time as Approach 2 took for games with 4 time steps.

## 6 Conclusions

TAGGs are a novel graphical representation of imperfect-information extensive-form games. They are an extension of simultaneous-move AGGs to the dynamic setting; and can be thought of as a sequence of AGGs played over $T$ time steps, with action counts accumulating as time progresses. This process can be formally described by the induced BN. For situations with anonymity or CSI structure, the TAGG representation can be exponentially more compact than a direct MAID representation. We presented an algorithm for computing expected utility for TAGGs that exploits its anonymity, CSI as well as temporal structure. We showed both theoretically and empirically that our approach is significantly more efficient than the standard approach on a direct MAID representation of the same game.